\documentclass[10pt]{iopart}

\usepackage[english]{babel}										
\usepackage[protrusion=true,expansion=true]{microtype}				
\usepackage[pdftex]{graphicx}									
\usepackage[svgnames]{xcolor}									
\usepackage[hang, small,labelfont=bf,up,textfont=it,up]{caption}	
\usepackage{epstopdf}												
\usepackage{subfig}													
\usepackage{booktabs}												
\usepackage{fix-cm}

\usepackage{txfonts}

\usepackage{amssymb}
\usepackage{txfonts}
\usepackage{color}
\usepackage{hyperref}
\hypersetup{
  colorlinks = true,
  citecolor = blue,
  urlcolor = black}

\newcommand{{\bfa}}{\mbox{\boldmath$a$\unboldmath}}
\newcommand{{\bfr}}{\mbox{\boldmath$r$\unboldmath}}
\newcommand{{\bfp}}{\mbox{\boldmath$p$\unboldmath}}
\newcommand{{\bfv}}{\mbox{\boldmath$v$\unboldmath}}
\newcommand{{\bff}}{\mbox{\boldmath$f$\unboldmath}}
\newcommand{{\bfF}}{\mbox{\boldmath$F$\unboldmath}}
\newcommand{{\bfA}}{\mbox{\boldmath$A$\unboldmath}}
\newcommand{{\bfchi}}{\mbox{\boldmath$\chi$\unboldmath}}

\newcommand{{\cF}}{\mbox{\boldmath${\cal F}$\unboldmath}}
\newcommand{{\cG}}{\mbox{\boldmath${\cal G}$\unboldmath}}
\newcommand{{\cE}}{\mbox{\boldmath${\cal E}$\unboldmath}}
\newcommand{{\cB}}{\mbox{\boldmath${\cal B}$\unboldmath}}
\newcommand{{\cX}}{\mbox{\boldmath${\cal X}$\unboldmath}}
\newcommand{{\cY}}{\mbox{\boldmath${\cal Y}$\unboldmath}}
\def\v#1{{\bf#1}}

\usepackage[top=2.5cm, bottom=2.5cm, left=3.0cm, right=3.0cm]{geometry}

\begin{document}

\title{\textcolor[rgb]{0.00,0.00,0.75}{\textsf An axiomatic approach to Maxwell's equations}}
\vskip 15pt

\author{Jos\'e A. Heras}

\address{Instituto de Geof\'isica, Universidad Nacional Aut\'onoma de M\'exico, Ciudad de M\'exico 04510, M\'exico}


\begin{indented}
\item[] E-mail: herasgomez@gmail.com
\end{indented}

\begin{abstract}
\noindent
This paper suggests an axiomatic approach to Maxwell's equations. The basis of this approach is a theorem formulated for two sets of functions localized in space and time. If each set satisfies a continuity equation then the theorem provides an integral representation for each function. A corollary of this theorem yields Maxwell's equations with magnetic monopoles. It is pointed out that the causality principle and the conservation of electric and magnetic charges are the most fundamental physical axioms underlying these equations. Another application of the corollary yields Maxwell's equations in material media. The theorem is also formulated in the Minkowski space-time and applied to obtain the covariant form of Maxwell's equations with magnetic monopoles and the covariant form of Maxwell's equations in material media. The approach makes use of the infinite-space Green function of the wave equation and is therefore suitable for an advanced course in electrodynamics.
\end{abstract}


\textcolor[rgb]{0.00,0.00,0.75}{\section{Introduction}}
\noindent Special relativity has been a very successful theory. A key part of its success is that it relies on a two relatively simple axioms: The relativity principle and the constancy of the speed of light $c.$\,\footnote[1]{These are the famous Einstein's original postulates of special relativity.  However, it can be reasonably argued that the constancy of $c$ (the second axiom) is a consequence of the relativity principle (the first axiom) together with natural symmetries of space-time, like homogeneity and isotropy. Many authors have derived generalizations of the Lorentz transformations involving an unspecified and invariant speed $k$ using only the first axiom and space-time symmetries [See, e.g., Levy-Leblond J M 1976 One more derivation of the Lorentz Transformation {\it Am. J. Phys.} \href{ http://dx.doi.org/10.1119/1.10490}{\textcolor[rgb]{0.00,0.00,1.00}{ {\bf 44}  271-77}}] In order to identify $k$ with $c$ we can invoke Maxwell's equations.} From these axioms we derive the Lorentz transformations which turn out to be vital for the construction and understanding of the theory itself. Classical electrodynamics has also been a very successful theory. We have also a set of equations, the famous Maxwell's equations, which are at the hearth of the theory. Nevertheless, the analogous ``simple'' axioms from which we can derive Maxwell's equations have not been clearly identified. This is particularly disturbing because it was precisely the form invariance of Maxwell's equations that played a central role in the derivation of the Lorentz transformations.\,\footnote[2]{A derivation of the Lorentz transformations from the invariance of the wave equation (Maxwell's equations can equivalently be represented by wave equations) has been recently presented by R. Heras 2016 Lorentz transformations and the wave
equation {\it Eur. J. Phys.} \href{ http://doi:10.1088/0143-0807/37/2/025603}{\textcolor[rgb]{0.00,0.00,1.00}{ {\bf 37} 025603}}].} In analogy with special relativity, it would be desirably to identify the two fundamental axioms of Maxwell's equations. Put differently, the profound link between classical electrodynamics and special relativity invites us to identify the fundamental physical axioms underlying Maxwell's equations. The student and the teacher would benefit from having a consistent axiomatic approach to Maxwell's equations as an alternative to standard approaches to these equations, which usually mix historical, heuristical and experimental elements.
Axiomatic presentations of quantum mechanics and general relativity have already appeared in textbooks \cite{1,2}. But axiomatic introductions to electrodynamics are not usually discussed in textbooks \cite{3}.

Any attempt to develop an axiomatic approach to Maxwell's equations must begin by identifying the fundamental physical axioms underlaying these equations. The problem is that there are several different postulates that can be classified as physical axioms for Maxwell's equations and therefore the task of recognizing which of them are the most fundamental is not so simple. Over the years, several authors have invoked fundamental symmetries to identify the physical axioms underlying Maxwell's equations. The story seems to have started with Page \cite{4} who
 derived electrodynamic equations from electrostatic equations using the Lorentz transformations. The derivation of Maxwell's equations from Coulomb's law and special relativity has  recurrently appeared in papers \cite{5,6,7} and textbooks \cite{8,9,10}.  This derivation has been criticized by       Jackson \cite{11} and Feynman \cite{12} who have pointed out that other non-trivial additional assumptions are required. The derivation of Maxwell's equations from the Biot-Savart law and special relativity has also been proposed \cite{13}. Gauge invariance is another symmetry that has been used to derive Maxwell's equations. Weyl \cite{14} was the first to derive Maxwell's equations using gauge arguments. A pedagogical version of this derivation was presented by Kobe \cite{15}. Another popular derivation of the homogeneous Maxwell equations, attributed to Feynman and presented by Dyson \cite{16} assumes Newton's law of motion and quantum mechanics commutation relations. Discrete symmetries like time reversal and spatial inversion have also been used to deduce Maxwell's equations \cite{17}.
Hehl and Obukhov \cite{3} have derived Maxwell's vacuum equations using four axioms: electric charge conservation, Lorentz force density, magnetic flux conservation and a constitutive law for the vacuum. We have pointed out that Maxwell's equations can be derived by starting with the continuity equation evaluated at the retarded time \cite{18}. We have claimed that charge conservation and causality are the most fundamental physical axioms underlying Maxwell's equations. We have also formulated this derivation of Maxwell's equations
in the Minkowski space-time \cite{19}.

In this paper we extend our axiomatic approach to Maxwell's equations. We show how Maxwell's equations with magnetic monopoles can be obtained by using the continuity equations for electric and magnetic charges and the infinity-space Green function of the wave equation. In section 2 we formulate a theorem for two sets of functions localized in space and time. Each set is formed by a scalar and a vector functions which satisfy a continuity equation. The theorem provides an integral representation for each function in terms of the derivatives of the involved functions and the retarded Green function. In section 3 we show how a corollary of the theorem provides an axiomatic derivation of Maxwell's equations with magnetic monopoles. The basic physical axioms in this derivation are the conservation of electric and magnetic charges represented by their associated continuity equations and the causality principle represented by the retarded Green function. In section 4 we apply again the corollary of the theorem to find Maxwell's equations in material media. In section 5 we formulate this theorem in  the Minkowski space-time and apply its associated corollary to obtain the covariant form of Maxwell's equations with magnetic monopoles and the covariant form of Maxwell's equations in material media. In section 6 we emphasize that charge conservation is a necessary but not a sufficient condition for Maxwell's standard equations and that causality is a sufficient but not a necessary condition for these equations. We also note the general validity of the corollary of the theorem and show how it can be applied to find the gravitational field equations proposed by Heaviside \cite{20} and discussed by Jefimenko \cite{21} and McDonald \cite{22}.

An extensive use of the infinite-space retarded Green function of the wave equation is made in the approach to Maxwell's equations suggested in this paper. As is well-known, this function is typically introduced in graduate textbooks of electrodynamics \cite{23,24}. Therefore the proposed approach is suitable for a graduate course of electrodynamics. However, in section 7 we outline an elementary introduction to this Green function which could be suitable for undergraduate students. If this Green function is introduced via this non-rigours and heuristic procedure then the three-dimensional version of the approach for Maxwell's equations proposed here, which involves well-known vector identities,  could be presented in an advanced undergraduate course of electrodynamics. In section 8 we present our general conclusion. Finally, in Appendices A and B we prove some useful identities.

\textcolor[rgb]{0.00,0.00,0.75}{\section{A theorem that leads to Maxwell's equations}}
\noindent {\textcolor[rgb]{0.00,0.00,0.75}{\textbf{Theorem.}}} Let $\rho_1\!=\!\rho_1(\v x,t)$ and $\rho_2\!=\!\rho_2(\v x,t)$ be two scalar functions of space and time. Let $\v J_1\!=\!\v J_1(\v x,t)$ and $\v J_2\!=\!\v J_2(\v x,t)$ be two vector functions of space and time. If these functions satisfy the continuity equations
\begin{eqnarray}
\nabla\cdot\v J_1+\frac{\partial\rho_1}{\partial t}=0,\\
\nabla\cdot\v J_2+\frac{\partial\rho_2}{\partial t}=0,
\end{eqnarray}
and are localized in space and time\footnote[3]{The functions $\rho_1, \rho_2, \v J_1$ and $\v J_2$ could also satisfy the condition that they vanish sufficiently rapidly at spatial infinity.}
then they admit the integro-differential  representations:
\begin{eqnarray}
\rho_1 = & \;\nabla\cdot \Bigg\{\frac{1}{4\pi}\!\iint \!\!d^3x'dt'G\Bigg(\!-\!\nabla'\rho_1-\frac{1}{\chi c^2}\bigg[\nabla'\times\v J_2+\chi\frac{\partial\v J_1}{\partial t'}\bigg]\Bigg)\Bigg\},\\
\rho_2 = &\;\nabla\cdot\Bigg\{\frac{1}{4\pi}\!\iint \!\!d^3x'dt'G\Bigg(\!\!-\!\nabla'\!\rho_2+\chi\bigg[\nabla'\times\v J_1-\frac{1}{\chi c^2}\frac{\partial\v J_2}{\partial t'}\bigg]\Bigg)\Bigg\},\\
\v J_1 =&\;\frac{1}{\chi} \!\nabla \!\times\Bigg\{\frac{1}{4\pi}\!\iint \!\!d^3x'dt'G\Bigg(\!\!-\nabla'\!\rho_2\!+\!\chi\bigg[\nabla'\times\v J_1\!-\!\frac{1}{\chi c^2}\frac{\partial\v J_2}{\partial t'}\bigg]\Bigg)\Bigg\}\nonumber\\&
-\!\frac{\partial}{\partial t}\Bigg\{\frac{1}{4\pi}
\iint \!\!d^3x'dt'G\Bigg(\!-\!\nabla'\!\rho_1\!-\!\frac{1}{\chi c^2}\bigg[\nabla'\times\v J_2\!+\!\chi\frac{\partial\v J_1}{\partial t'}\bigg]\Bigg)\Bigg\},\\
\v J_2 =&-\chi c^2\nabla\!\times\!\Bigg\{\frac{1}{4\pi}\!\iint \!\!d^3x'dt'G\Bigg(\!\!-\!\nabla'\!\rho_1\!-\!\frac{1}{\chi c^2}\bigg[\nabla'\times\v J_2\!+\!\chi\frac{\partial\v J_1}{\partial t'}\bigg]\Bigg)\Bigg\} \nonumber\\&-\!\frac{\partial}{\partial t}\Bigg\{\frac{1}{4\pi} \iint \!\!d^3x'dt'G\Bigg(\!\!-\!\nabla'\!\rho_2\!+\!\chi\bigg[\nabla'\times\v J_1\!-\!\frac{1}{\chi c^2}\frac{\partial\v J_2}{\partial t'}\bigg]\Bigg)\Bigg\},
\end{eqnarray}
where $G\!\equiv\! G(\v x,t;\v x',t')$ is the infinity-space Green function of the wave equation and $\chi$ and  $c$ are positive constants having the latter dimensions of velocity.\,\footnote[6]{The system formed by equations (1)-(2) is undetermined. It involves two scalar equations with eight unknowns! Informally speaking, however, we can say that equations (3)-(6) constitute a ``solution'' for equations (1) and (2). In a more formal sense, equations (3)-(6) are integro-differential representations of the functions $\rho_1, \rho_2, \v J_1$ and $\v J_2$ (expressed in terms of their spatial and temporal derivatives) which are seen to satisfy equations (1) and (2).}

 \noindent  {\textcolor[rgb]{0.00,0.00,0.75}{\bf Corollary.}} If the vector fields $\v F_1\!=\!\v F_1(\v x,t)$ and $\v F_2\!=\!\v F_2(\v x,t)$ are defined by\,\footnote[7]{Following a terminology typical of field theory, we observe that the ``sources'' $\rho_1$ and $\v J_1$ in equation (1) are independent from the ``sources'' $\rho_2$ and $\v J_2$ in equation (2). However, the ``field'' $\v F_1$ in equation (7) connects $\rho_1$ with $\v J_1$ and $\v J_2$  and the ``field'' $\v F_2$ in equation (8) relates $\rho_2$ with $\v J_1$ and $\v J_2$. In other words, the fields in equations (7) and (8) couple the sources present in equations (1) and (2). Since equations (1) and (2) can represent physical symmetries (conservation laws) then
 $\v F_1$ and $\v F_1$ can be physically interpreted as manifestations of these symmetries.}
\begin{eqnarray}
\v F_1 &= \frac{1}{4\pi}\!\iint \!\!d^3x'dt'G\Bigg(\!\!-\!\nabla'\!\rho_1\!-\!\frac{1}{\chi c^2}\bigg[\nabla'\times\v J_2+\chi\frac{\partial\v J_1}{\partial t'}\bigg]\Bigg),\\
\v F_2 &=  \frac{1}{4\pi}\!\iint \!\!d^3x'dt'G\Bigg(\!\!-\!\nabla'\!\rho_2+\chi\bigg[\nabla'\times\v J_1-\frac{1}{\chi c^2}\frac{\partial\v J_2}{\partial t'}\bigg]\Bigg),
\end{eqnarray}
then equations (3)-(6) take the compact form\,\footnote[9]{Once the functions $\rho_1,\, \rho_2, \,\v J_1$ and $\v J_2$ are specified, the system formed by equations (9)-(12) describes eight scalar equations with six unknowns. At first glance, this system seems to be over-determined because there are more equations than unknowns. However, this is not so because we have two additional restrictions: $\nabla\cdot [(1/\chi)\nabla\times \v F_2-\partial\v F_1/\partial t] +\partial[\nabla\cdot\v F_1]/\partial t\equiv 0$ and $\nabla\cdot\v [-\chi c^2\nabla\times \v F_1-\partial\v F_2/\partial t]+\partial[\nabla\cdot\v F_2]/\partial t\equiv 0$ which correspond to $\nabla\cdot\v J_1+\partial\rho_1/\partial t=0$ and $\nabla\cdot\v J_2+\partial\rho_2/\partial t=0$. The system in equations (9)-(12) is determined (given appropriate boundary conditions).}
\begin{eqnarray}
\qquad\qquad\nabla\cdot\v F_1=\rho_1,\\
\qquad\qquad\nabla\cdot\v F_2=\rho_2,\\
\quad\frac{1}{\chi}\nabla\times \v F_2-\frac{\partial\v F_1}{\partial t}=\v J_1,\\
-\chi c^2\nabla\times \v F_1-\frac{\partial\v F_2}{\partial t}= \v J_2.
\end{eqnarray}
The reader can admire from now on the similarity of equations (9)-(12) with respect to Maxwell's equations with electric and magnetic charges.
As a first step to show the theorem, we will first see that equations (1) and (2) are form invariant under the duality transformations:
\begin{eqnarray}
\rho_1'&= \rho_1 \cos\theta +\frac{1}{\chi c}\rho_2 \sin \theta, \\
\rho_2'&= -\chi c \rho_1 \sin \theta + \rho_2 \cos \theta,\\
\v J_1'&= \v J_1 \cos\theta +\frac{1}{\chi c}\v J_2 \sin \theta,\\
\v J_2'&= -\chi c \v J_1\sin \theta + \v J_2 \cos \theta,
\end{eqnarray}
where $\theta$ is a real angle. The proof of this invariance is straightforward
\begin{eqnarray}
\nabla\cdot\v J_1'= \nabla\cdot\v J_1 \cos\theta +\frac{1}{\chi c}\nabla\cdot\v J_2 \sin \theta
=-\frac{\partial\rho_1}{\partial t} \cos\theta -\frac{1}{\chi c}\frac{\partial\rho_2}{\partial t} \sin \theta\nonumber\\
\qquad =-\frac{\partial}{\partial t}\bigg(\rho_1\cos\theta -\frac{1}{\chi c}\rho_2 \sin \theta\bigg)
=-\frac{\partial\rho_1'}{\partial t},\\
\nabla\cdot\v J_2'= -\chi c \nabla\cdot\v J_1\sin \theta + \nabla\cdot\v J_2 \cos \theta
=\chi c\frac{\partial\rho_1}{\partial t}\sin \theta - \frac{\partial\rho_2}{\partial t}\cos \theta\nonumber\\
\qquad =-\frac{\partial}{\partial t}\big(\!\!-\chi c\rho_1\sin \theta +\rho_2\cos \theta\big)
=-\frac{\partial\rho_2'}{\partial t},
\end{eqnarray}
where we have used equations (1), (2) and (13)-(16). For $\theta=\pi/2$ equations (13)-(16) imply
\begin{eqnarray}
\rho_1\rightarrow \frac{1}{\chi c}\rho_2, \quad
 \rho_2 \rightarrow-\chi c \rho_1,\quad
 \v J_1 \rightarrow\frac{1}{\chi c}\v J_2\quad
 \v J_2\rightarrow -\chi c \v J_1,
\end{eqnarray}
If these changes are applied to  equation
(1) then we obtain equation (2). Viceversa, when the changes given in equation (19) are applied to
equation (2) we obtain equation (1). This dual property generally extends to equations derived from equations (1) and (2).

As a second step, we will show that equations (1) and (2) imply the following identities:
\begin{eqnarray}
\!\!\!\!\!\!\!\!\!\!\!\!\!\!\!\!\!\!\!\!\!\!\!\!\!\!\!\!\!\!\!\!\!\!\!\!\!\!\Box^{2}\rho_1 = -\nabla\cdot\Bigg(\!-\!\nabla\rho_1-\frac{1}{\chi c^2}\bigg[\nabla\times\v J_2+\chi \frac{\partial\v J_1}{\partial t}
\bigg]\Bigg),\\
\!\!\!\!\!\!\!\!\!\!\!\!\!\!\!\!\!\!\!\!\!\!\!\!\!\!\!\!\!\!\!\!\!\!\!\!\!\!\Box^{2}\rho_2 = -\nabla\cdot\Bigg(\!-\!\nabla\rho_2 +\chi\bigg[\nabla\times\v J_1-\frac{1}{\chi c^2}\frac{\partial\v J_2}{\partial t}\bigg]\Bigg), \\
\!\!\!\!\!\!\!\!\!\!\!\!\!\!\!\!\!\!\!\!\!\!\!\!\!\!\!\!\!\!\!\!\!\!\!\!\!\!\Box^{2}\v J_1 = -\frac{1}{\chi}\nabla\times \Bigg(\!-\nabla\rho_2+\chi\bigg[\nabla\times\v J_1-\frac{1}{\chi c^2}\frac{\partial\v J_2}{\partial t}\bigg]\Bigg)+\frac{\partial}{\partial t}\Bigg(\!-\!\nabla\rho_1-\frac{1}{\chi c^2}\bigg[\nabla\times\v J_2+\chi\frac{\partial\v J_1}{\partial t}\bigg]\Bigg),\\
\!\!\!\!\!\!\!\!\!\!\!\!\!\!\!\!\!\!\!\!\!\!\!\!\!\!\!\!\!\!\!\!\!\!\!\!\!\!\Box^{2}\v J_2= \chi c^2\nabla\times \Bigg(\!-\nabla\rho_1-\frac{1}{\chi c^2}\bigg[\nabla\times\v J_2+\chi\frac{\partial\v J_1}{\partial t}\bigg]
\Bigg)+\frac{\partial}{\partial t}\Bigg(\!-\!\nabla\rho_2+\chi\bigg[\nabla\times\v J_{\rm 1}-\frac{1}{\chi c^2}\frac{\partial\v J_{\rm 2}}{\partial t}\bigg]\Bigg),
\end{eqnarray}
where $\Box^{2}\equiv\nabla^2-(1/c^2)\partial^2/\partial t^2$ is the d'Alembert operator.

We apply $-(1/c^2) \partial/\partial t$ to equation (1) and obtain $-(1/c^2)\partial^2 \rho_1/\partial t^2=(1/c^2)\partial (\nabla\cdot\v J_1)/\partial t.$ In the right-hand side we add the identically zero term $[1/(\chi c^2)]\nabla\cdot(\nabla\times\v J_2)\equiv0$. We obtain
\begin{eqnarray}
-\frac{1}{c^2}\frac{\partial^2\rho_1}{\partial t^2}=\nabla\cdot\Bigg(\frac{1}{\chi c^2}\bigg[\nabla\times\v J_2+ \chi\frac{\partial\v J_1}{\partial t}\bigg]\Bigg).
\end{eqnarray}
We add $\nabla^2\rho_1$ to both sides of equation (24) and, after factoring, we get equation (20).

Now we apply $\nabla$  to equation (1) and obtain $\nabla(\nabla\cdot\v J_1)\!=\!-\partial\nabla\rho_1/\partial t$, which implies $\nabla^2\v J_1\!=\!-\nabla\times(\nabla\times \v J_1)\! -\!\nabla\partial\rho_1/\partial t.$ In the right-hand side of this equation we
add the identically zero term  $-(1/\chi)\nabla\times \nabla\rho_2\!-\![1/(\chi c^2)]\nabla\times\partial\v J_2/\partial t\! + \![1/(\chi c^2)]\partial(\nabla\times\v J_2)/\partial t\equiv 0$. We obtain
\begin{eqnarray}
\!\!\!\!\!\!\!\!\!\!\!\!\!\!\!\!\!\!\!\!\nabla^{2}\v J_1= -\frac{1}{\chi}\nabla\times \Bigg(\!-\nabla\rho_2+\chi\bigg[\nabla\times\v J_1-\frac{1}{\chi c^2}\frac{\partial\v J_2}{\partial t}\bigg]\Bigg)+\frac{\partial}{\partial t}\Bigg(\!-\!\nabla\rho_1-\frac{1}{\chi c^2}\bigg[\nabla\times\v J_2\bigg]\Bigg).\qquad
\end{eqnarray}
We add $-(1/c^2)\partial^2\v J_1/\partial t^2$ to both sides of equation (25) and obtain equation (22). The remaining equations (21) and (23) are obtained by applying the dual changes displayed in equation (19) to equations (20) and (22).

We will use the infinity-space Green function of the wave equation \cite{23} as an integrating factor for equations (20)-(23) with the purpose of obtaining  integral representations of the four functions appearing in equations (3)-(6). This Green function is commonly used to solve Maxwell's equations \cite{25,26}. It satisfies the wave equation $\Box^{2}G= -4\pi\delta(\v x-\v x')\delta(t-t')$, where $\delta$ is the Dirac delta function, and the properties $\nabla G =-\nabla' G$ and $\partial G/\partial t=-\partial G/\partial t'.$ Using these properties, the following results are proved in the Appendix A:
\begin{eqnarray}
\;\iint \!\!d^3x'dt'G\nabla'\diamond {\cal F}=\nabla\diamond\iint \!\!d^3x'dt'G{\cal F},\\
\quad\;\,\iint \!\!d^3x'dt'G\frac{\partial {\cal F}}{\partial t'}=\frac{\partial}{\partial t}\iint \!\!d^3x'dt'G{\cal F},\\
\;\;\,\iint \!\!d^3x'dt'G\Box'^2{\cal F} =\iint \!\!d^3x'dt'{\cal F}\Box'^2G\nonumber\\
\qquad\qquad \qquad\;\;\; =-4\pi{\cal F}(\v x,t),
\end{eqnarray}
where the spatial integration is over all space and the time integration is over all time from $-\infty$ to $+\infty$.
The symbol $\diamond$ represents a generic product that may be the dot product, the cross product and the direct product. The letter ${\cal F}$ denotes a well-behaved function of space and time that may represent a scalar function
, a vector function and a tensor function.  For example, if ${\cal F}=f(\v x,t)$ is a scalar function then $\nabla\diamond {\cal F}$ is the gradient $\nabla\diamond {\cal F}=\nabla f$.
If ${\cal F}=\v F(\v x,t)$ is a vector function then $\nabla\diamond {\cal F}$ can represent either the dot product $\nabla\diamond {\cal F}=\nabla\cdot \v F$ or the cross product $\nabla\diamond {\cal F}=\nabla\times \v F$. The function ${\cal F}$ is assumed to be localized in space and time.

If we evaluate equation (20) at the point $\v x'$ and the time $t'$, multiply the resulting equation by the Green function $G$ and
integrate over all space and all time, then we obtain the equation
\begin{eqnarray}
\!\!\!\!\!\!\!\!\!\!\!\!\iint \!\!d^3x'dt'G\Box'^{2}\rho_1=-\iint \!\!d^3x'dt'G\nabla'\cdot\Bigg(\!-\!\nabla'\rho_1-\frac{1}{\chi c^2}\bigg[\nabla'\times\v J_2+\chi \frac{\partial\v J_1}{\partial t'}
\bigg]\Bigg).
\end{eqnarray}
When equations (26) and (28) are used in equation (29), we find equation (3). Similarly, we evaluate equation (22) at the point $\v x'$ at the time $t'$, multiply the resulting equation by $G$ and
integrate over all space and all time:
\begin{eqnarray}
\!\!\!\!\!\!\!\!\!\!\!\!\iint \!\!d^3x'dt'G\Box'^{2}\v J_1=&\iint \!\!d^3x'dt'G\Bigg\{\!-\frac{1}{\chi}\nabla'\times \Bigg(\!\nabla'\rho_2\!+\!\chi\bigg[\nabla'\times\v J_1\!-\frac{1}{\chi c^2}\frac{\partial\v J_2}{\partial t'}\bigg]\Bigg)\nonumber\\
& +\!\frac{\partial}{\partial t'}\Bigg(\!-\!\nabla'\rho_1\!-\!\frac{1}{\chi c^2}\bigg[\nabla'\!\times\!\v J_2\!+\!\chi\frac{\partial\v J_1}{\partial t'}\bigg]\Bigg)\Bigg\}.
\end{eqnarray}
When equations (26)-(28) are used in equation (30) we find equation (5). Equations~(3) and (6) follow from applying the changes defined in equation (19) to equations (4) and (5). Once equations (3)-(6) have been obtained, the theorem has been demonstrated.

The corollary expressed in equations (7)-(12) can be easily proved. We simply substitute equations (7) and (8) into equations (3)-(6). We observe that the set formed by equations (7)-(12) is invariant under the set formed by the dual transformations defined in equations (13)-(16) together with the dual transformations
\begin{eqnarray}
\v F_1'&= \v F_1 \cos\theta +\frac{1}{\chi c}\v F_2 \sin \theta,\\
\v F_2'&= -\chi c \v F_1\sin \theta + \v F_2 \cos \theta.
\end{eqnarray}

Integration of the wave equation $\Box^{2}G\!=\! -4\pi\delta(\v x-\v x')\delta(t-t')$
yields two physically-different solutions \cite{23, 24}: $G^{(\pm)}\!= \!\delta(t'\!-[t\mp R/c])/R$. The retarded solution $G^{(+)}=\delta(t'-t+ R/c)/R $ exhibits the causal behaviour associated with the wave disturbance. The advanced Green function $G^{(-)}\!=\!\delta(t'\!-t\!- \!R/c)/R $ is associated with a non-causal behavior of the wave disturbance. For physical applications we must identify $G$ with the retarded form $G^{(+)}$ to be consistent with the causality principle. We also note that equations (3)-(6) can alternatively be justified by simply taking their d'Alembertian and showing that four identities result. In fact, if we apply the d'Alembertian to equations (3)-(6) and perform the corresponding integrations over all space and all time then we obtain the identities appearing in equations (20)-(23).\\

\textcolor[rgb]{0.00,0.00,0.75}{\section{Maxwell's equations with magnetic monopoles}}

\noindent The theorem proved in section 2 can be applied to Maxwell's equations provided $c$ is identified with the speed of light in vacuum and the constants $\alpha, \beta$ and $\chi$ are related by \cite{18,27}: $ \alpha=\beta\chi c^2$. Specific values of these constants corresponding to Gaussian units, International System of Units (SI) and Heaviside-Lorentz (H-L) units appear in the table 1.
\begin{table}[h]
\qquad\qquad \qquad\;\begin{tabular}{|l|l|l|l|}
\hline
 System & \;\,$\alpha$ &\;\, $\beta\;$ \;&\; $\chi\;\;$\\
\hline
Gaussian &\, $4\pi$ & $4\pi/c $ & $1/c$\\ \hline
SI & $1/\epsilon_0$&$\;\;\mu_0$&\; 1\\ \hline
Heaviside-Lorentz &\; $1$ & $\;1/c $ & $1/c$\\
\hline
\end{tabular}
\caption{Specific values of $\alpha, \beta$ and $\chi$ for three unit systems }
\end{table}

In a first application of the corollary of the theorem, we will derive Maxwell's equations with magnetic monopoles. We make the following identifications:
\begin{eqnarray}
\v F_1=\frac{1}{\alpha}\v E,\;\; \v F_2=\frac{\chi}{\beta} \v B, \;\; \rho_1=\rho_{\rm e},\;\;  \rho_2=\rho_{\rm m},\;\;  \v J_1=\v J_{\rm e},\;\;  \v J_2=\v J_{\rm m},
\end{eqnarray}
where $\rho_{\rm e}$ and $\v J_{\rm e}$ are the electric charge and current densities and $\rho_m$ and $\v J_m$ are the magnetic charge and current densities. The densities $\rho_{\rm e}$ and $\v J_{\rm e}$ are localized in space and time and satisfy the continuity equation $\nabla\cdot\v J_{\rm e}+ \partial\rho_{\rm e}/\partial t=0.$ Similarly, the densities $\rho_{\rm m}$ and $\v J_{\rm m}$ are localized in space and time and satisfy the
continuity equation: $\nabla\cdot\v J_{\rm m}+ \partial\rho_{\rm m}/\partial t=0.$ With the identifications in equation (33), we can see that equations (7) and (8) become the time-dependent electric and magnetic fields produced by electric and magnetic sources in vacuum:
\begin{eqnarray}
\v E= &\;\frac{\alpha}{4\pi}\iint \!\!d^3x'dt'G\Bigg(\!\!-\!\nabla'\!\rho_{\rm e}\!-\!\frac{1}{\chi c^2}\bigg[\nabla'\!\times\!\v J_{\rm m}\!+\!\chi\frac{\partial\v J_{\rm e}}{\partial t'}\bigg]\Bigg),\\
\v B= & \;\frac{\beta}{4\pi\chi}\iint \!\!d^3x'dt'G\Bigg(\!\!-\!\nabla'\!\rho_{\rm m}+\chi\bigg[\nabla'\times\v J_{\rm e}-\frac{1}{\chi c^2}\frac{\partial\v J_{\rm m}}{\partial t'}\bigg]\Bigg),
\end{eqnarray}
and equations (9)-(12) become Maxwell's equations with magnetic monopoles \cite{27}:
\begin{eqnarray}
 \qquad\quad\nabla\cdot\v E=\alpha\rho_{\rm e},\\
\qquad\quad\nabla\cdot\v B=\frac{\beta}{\chi}\rho_{\rm m},\\
\nabla\times \v B-\frac{\beta}{\alpha}\frac{\partial\v E}{\partial t}=\beta\v J_{\rm e},\\
\;\nabla\times \v E+\chi\frac{\partial\v B}{\partial t} = -\beta\v J_{\rm m},
\end{eqnarray}
which are expressed in a form independent of specific units. Some comments are in line:
\begin{enumerate}
  \item As already noted, in order to have physical fields the function $G$ in equations (34) and (35) must be identified with the retarded Green function of the wave equation.
  \item  Time integration in equations (34) and (35) yields the more familiar retarded forms:
\begin{eqnarray}
\v E= &\frac{\alpha}{4\pi}\!\int d^3x'\frac{1}{R}\Bigg[\!-\!\nabla'\!\rho_e\!-\!\frac{1}{\chi c^2}\Bigg(\nabla'\!\times\!\v J_{\rm m}\!+\!\chi\frac{\partial\v J_{\rm e}}{\partial t'}\Bigg)\Bigg]_{\rm ret},\\
\v B= & \frac{\beta}{4\pi\chi} \!\int d^3x'\frac{1}{R}\Bigg[\!-\!\nabla'\!\rho_{\rm m}+\chi\Bigg(\nabla'\times\v J_{\rm e}-\frac{1}{\chi c^2}\frac{\partial\v J_{\rm m}}{\partial t'}\Bigg)\Bigg]_{\rm ret},
\end{eqnarray}
where the square bracket $[\quad]_{\rm ret}$ means that the enclosed quantity is to be evaluated at the retarded time $t'=t-R/c$.
  \item In absence of magnetic charges, equations (34) and (35) reduce to
\begin{eqnarray}
\v E &= \frac{\alpha}{4\pi}\!\iint \!\!d^3x'dt'G\Bigg(\!-\!\nabla'\!\rho_{\rm e}\!-\!\frac{1}{c^2}\frac{\partial\v J_{\rm e}}{\partial t'}\Bigg),\\
\v B &= \frac{\beta}{4\pi}\!\iint \!\!d^3x'dt'G\nabla'\times\v J_{\rm e},
\end{eqnarray}
or equivalently,
\begin{eqnarray}
\v E&=\frac{\alpha}{4\pi}\!\int d^3x'\frac{1}{R}\Bigg[\!-\!\nabla'\!\rho_{\rm e}\!-\!\frac{1}{c^2}\frac{\partial\v J_{\rm e}}{\partial t'}\Bigg]_{\rm ret},\\
\v B&= \frac{\beta}{4\pi} \!\int d^3x'\frac{1}{R}\big[\nabla'\times\v J_{\rm e}\big]_{\rm ret}.
\end{eqnarray}
and equations (36)-(39) reduce to Maxwell's standard equations \cite{18}:
\begin{eqnarray}
 \qquad\quad\nabla\cdot\v E=\alpha\rho_{\rm e},\\
 \qquad\quad\nabla\cdot\v B=0,\\
\nabla\times \v B-\frac{\beta}{\alpha}\frac{\partial\v E}{\partial t}=\beta\v J_{\rm e},\\
\;\nabla\times \v E+\chi\frac{\partial\v B}{\partial t} = 0.
\end{eqnarray}
  \item Using the identifications in equation (33), equations (13)-(16), (31) and (32) become
\begin{eqnarray}
\rho_{\rm e}'&= \rho_{\rm e} \cos\theta +\frac{1}{\chi c}\rho_{\rm m} \sin \theta,\\
 \rho_{\rm m}'&= -\chi c \rho_e \sin \theta + \rho_{\rm m} \cos \theta,\\
 \v J_{\rm e}'&= \v J_{\rm e}\cos\theta +\frac{1}{\chi c}\v J_{\rm m} \sin \theta,\\
 \v J_{\rm m}'&= -\chi c \v J_{\rm e} \sin \theta + \v J_{\rm m} \cos \theta,\\
 \v E'&= \v E \cos\theta +\chi c\v B \sin \theta,\\
\v B'&= -\frac{1}{\chi c}\v E \sin\theta +\v B \cos \theta.
\end{eqnarray}
\end{enumerate}
These are the duality transformations of Maxwell's equations with magnetic monopoles.
\textcolor[rgb]{0.00,0.00,0.75}{\section{Maxwell's equations in material media}}
\noindent In the second application of the corollary, we make the following identifications:
\begin{eqnarray}
\!\!\!\!\!\!\!\!\!\!\!\!\!\!\!\!\!\!\!\!\!\!\!\!\!\!\!\!\!\v F_1=\frac{1}{\alpha}\v E,\; \v F_2=\frac{\chi}{\beta} \v B, \;\rho_1=\rho_{\rm e}-\nabla\cdot\v P,\;\;  \rho_2=0,\;  \v J_1=\v J_{\rm e}+\frac{\partial\v P}{\partial t} + \frac{1}{\chi}\nabla\times \v M,\;  \v J_2=0,
\end{eqnarray}
where $-\nabla\!\cdot\!\v P$ and $\partial\v P/\partial t\! +\! \nabla\!\times\!\v M/\chi$ represent bound charge and current densities
with $\v P$ and $\v M$ being the polarization and magnetization vectors. Using the identifications given in equation (56), it follows that equations (7) and (8) yield the electric and magnetic fields produced by free and bound sources:
\begin{eqnarray}
\!\!\!\!\!\!\!\!\!\v E= &\;\frac{\alpha}{4\pi}\iint \!\!d^3x'dt'G\Bigg(\!\!-\!\nabla'(\rho_{\rm e}-\nabla'\cdot\v P)  \!-\!\frac{1}{c^2}
\frac{\partial}{\partial t'}\bigg[\v J_{\rm e}+\frac{\partial\v P}{\partial t'} + \frac{1}{\chi}\nabla'\times \v M\bigg]\Bigg),\\
\!\!\!\!\!\!\!\!\!\v B= & \;\frac{\beta}{4\pi}\iint \!\!d^3x'dt'G \nabla'\times\bigg(\v J_{\rm e}+\frac{\partial\v P}{\partial t'} + \frac{1}{\chi}\nabla'\times \v M\bigg),
\end{eqnarray}
or equivalently,
\begin{eqnarray}
\v E= &\;\frac{\alpha}{4\pi}\!\int d^3x'\Bigg[\!-\!\nabla'(\rho_e-\nabla'\cdot\v P)  \!-\!\frac{1}{c^2}
\frac{\partial}{\partial t'}\bigg(\v J_{\rm e}+\frac{\partial\v P}{\partial t'} + \frac{1}{\chi}\nabla'\times \v M\bigg)\Bigg]_{\rm ret},\\
\v B= & \;\frac{\beta}{4\pi}\!\int d^3x' \nabla'\times\Bigg[\v J_{\rm e}+\frac{\partial\v P}{\partial t'} + \frac{1}{\chi}\nabla'\times \v M\Bigg]_{\rm ret},
\end{eqnarray}
and equations (9)-(12) become Maxwell's equations in material media \cite{27}:
\begin{eqnarray}
 \qquad\quad\nabla\cdot\v E=\alpha\big(\rho_{\rm e}-\nabla\cdot\v P\big),\\
 \qquad\quad\nabla\cdot\v B=0,\\
\nabla\times \v B-\frac{\beta}{\alpha}\frac{\partial\v E}{\partial t}=\beta\Bigg( \v J_{\rm e}+\frac{\partial\v P}{\partial t} + \frac{1}{\chi}\nabla\times \v M      \Bigg),\\
\;\nabla\times \v E+\chi\frac{\partial\v B}{\partial t} = 0,
\end{eqnarray}
If we define the vector fields $\v D$ and $\v H$ as
\begin{eqnarray}
\v D &\equiv k_1\big(\v E+\alpha\v P\big),\\
\v H &\equiv k_2\bigg(\v B-\frac{\beta}{\chi}\v M\bigg),
\end{eqnarray}
then equations (61) and (63) can alternatively be written as \cite{27}:
\begin{eqnarray}
\qquad\qquad\nabla\cdot \v D=k_1\alpha\rho_{\rm e},\\
\nabla\times \v H-\frac{k_2}{k_1}\frac{\beta}{\alpha}\frac{\partial \v D}{\partial t}= k_2\beta\v J_{\rm e}.
\end{eqnarray}
For suitable values of the constants $k_1$ and $k_2$, equations (65)-(68) yield their corresponding specialized forms for Gaussian, SI and HL units. Table 2 shows the values of $k_1$ and $k_2$.
\begin{table}[h]
\qquad\qquad \quad\;\;\; \begin{tabular}{|l|l|l|l|}
\hline
System &\, $k_1$ &\; $k_2$\\
\hline
Gaussian & \;1 &\; 1  \\ \hline
SI & $\,\epsilon_0$&$1/\mu_0$\\ \hline
Heaviside-Lorentz &\;1 &\; 1 \\
\hline
\end{tabular}
\caption{Values of the constants $k_1$ and $k_2$.}
\end{table}

\vskip 10pt
\textcolor[rgb]{0.00,0.00,0.75}{\section{The covariant treatment}}

\noindent The theorem proved in section 2 can be formulated in the Minkowski space-time. Greek indices $\mu, \nu, \kappa \ldots$ run from 0 to 3 and  Latin indices $i,j,k,\ldots$ run from 1 to 3. The summation convention on repeated indices is adopted. The signature of the metric is $(+,-,-,-).$ A point is denoted by $x=x^{\mu}=\big(x^0,\v x\big)=\big(ct,\v x\big).$ The four gradient is defined as $\partial_\mu=\big[(1/c)\partial/\partial t, \nabla\big].$ The totally antisymmetric four-dimensional tensor reads $\varepsilon^{\mu\nu\alpha\beta}$ with $\varepsilon^{0123}=1$ and $\varepsilon^{ijk}$ is the totally antisymmetric three-dimensional tensor with $\varepsilon^{123} = 1$.
\vskip 5pt
\noindent  \textcolor[rgb]{0.00,0.00,0.75}{\textbf{Theorem.}} If the four-vectors $J^\nu_{\{1\}}\!=\!J^\nu_{\{1\}}(x)$ and $J^\nu_{\{2\}}\!=\!J^\nu_{\{2\}}(x)$ are localized in space-time and satisfy the continuity equations
\begin{eqnarray}
\partial_\nu J^\nu_{\{1\}} =0,\quad
\partial_\nu J^\nu_{\{2\}} =0,
\end{eqnarray}
then they admit the following integral representations:
\begin{eqnarray}
J^\nu_{\{1\}}&= \partial_\mu\! \int\! d^4x' D\,\Bigg(\partial'^\mu J^\nu_{\{1\}}-\partial'^\nu J^\mu_{\{1\}}-\frac{1}{\chi
 c}\varepsilon^{\mu\nu\kappa\lambda}\partial'_\kappa J_\lambda^{\{2\}}\Bigg),\\
J^\nu_{\{2\}}&=\chi c \,\partial_\mu\! \int\! d^4x' D\,\Bigg( \frac{1}{\chi c}\big(\partial'^\mu J^\nu_{\{2\}}-\partial'^\nu J^\mu_{\{2\}}\big)+ \varepsilon^{\mu\nu\kappa\lambda}\partial'_\kappa J_\lambda^{\{1\}}\Bigg),
\end{eqnarray}
where $D$ is the infinity-space Green function of the four-dimensional wave equation and the integration is over all space-time.
\vskip 5pt
 \noindent \textcolor[rgb]{0.00,0.00,0.75}{\textbf{Corollary.}} If the tensor fields $F^{\mu\nu}_{\{1\}}=F^{\mu\nu}_{\{1\}}(x)$ and $F^{\mu\nu}_{\{2\}}=F^{\mu\nu}_{\{2\}}(x)$ are defined as
\begin{eqnarray}
F^{\mu\nu}_{\{1\}}&=\int\! d^4x' D\,\Bigg(\partial'^\mu J^\nu_{\{1\}}-\partial'^\nu J^\mu_{\{1\}}-\frac{1}{\chi
 c}\varepsilon^{\mu\nu\kappa\lambda}\partial'_\kappa J_\lambda^{\{2\}}\Bigg),\\
F^{\mu\nu}_{\{2\}}&=\int\! d^4x' D\,\Bigg( \frac{1}{\chi c}\big(\partial'^\mu J^\nu_{\{2\}}-\partial'^\nu J^\mu_{\{2\}}\big)+ \varepsilon^{\mu\nu\kappa\lambda}\partial'_\kappa J_\lambda^{\{1\}}\Bigg),
\end{eqnarray}
then equations (70) and (71) take the compact form
\begin{eqnarray}
 \partial_\mu F^{\mu\nu}_{\{1\}}=J^\nu_{\{1\}},\quad
\chi c \partial_\mu F^{\mu\nu}_{\{2\}}=J^\nu_{\{2\}}.
\end{eqnarray}

Before proving the theorem, we observe that the continuity equations given in (69) are invariant under the dual transformations:
\begin{eqnarray}
J'^\nu_{\{1\}}&=J^\nu_{\{1\}} \cos\theta +\frac{1}{\chi c}J^\nu_{\{2\}}\sin \theta, \\
J'^\nu_{\{2\}}&= -\chi c J^\nu_{\{1\}}\sin \theta + J^\nu_{\{2\}}\cos \theta,
\end{eqnarray}
where $\theta$ is a real angle. We can directly verify this invariance
\begin{eqnarray}
\partial_\nu J'^\nu_{\{1\}}&=\partial_\nu J^\nu_{\{1\}} \cos\theta +\frac{1}{\chi c} \partial_\nu J^\nu_{\{2\}}\sin \theta=0, \\
\partial_\nu J'^\nu_{\{2\}}&= -\chi c \partial_\nu J^\nu_{\{1\}}\sin \theta + \partial_\nu J^\nu_{\{2\}}\cos \theta=0,
\end{eqnarray}
where we have used equations (69), (75) and (76). In particular, the continuity equations defined in equation (69) are shown to be invariant under the specific dual transformations:
\begin{eqnarray}
 J^\nu_{\{1\}} \rightarrow\frac{1}{\chi c}J^\nu_{\{2\}}, \quad
 J^\nu_{\{2\}} \rightarrow -\chi c J^\nu_{\{1\}},
\end{eqnarray}
which follow from equations (75) and (76) when $\theta=\pi/2$. Accordingly, when the dual changes given in equations (79) are applied to some result derived from the first continuity equation appearing in equation (69), we obtain the analogous result corresponding to the second continuity equation  appearing in equation (69). For example, when the dual changes defined in equation (79) are applied to equation (70), we obtain equation (71).

The continuity equations given in equation (69) imply the following identities
\begin{eqnarray}
\partial_\mu\partial^\mu J^\nu_{\{1\}}&= \partial_\mu\Bigg(\partial^\mu J^\nu_{\{1\}}-\partial^\nu J^\mu_{\{1\}}-\frac{1}{\chi c} \varepsilon^{\mu\nu\kappa\lambda}\partial_\kappa J_\lambda^{\{2\}}\Bigg),\\
\partial_\mu\partial^\mu J^\nu_{\{2\}}&= \chi c\,
\partial_\mu \Bigg(\frac{1}{\chi c}\big(\partial^\mu J^\nu_{\{2\}}-\partial^\nu J^\mu_{\{2\}}\big)+\varepsilon^{\mu\nu\kappa\lambda}\partial_\kappa J_\lambda^{\{1\}}\Bigg).
\end{eqnarray}
To construct equation (80) we apply $\partial^\mu$ to the first continuity equation appearing in equation (69) obtaining $\partial^\mu\partial_\nu J^\nu_{\{1\}} =0$.
 We add the term $\partial_\mu\partial^\mu J^\nu_{\{1\}}$ in both sides of the last equation and obtain $\partial_\mu\partial^\mu J^\nu_{\{1\}}= \partial_\mu\big(\partial^\mu J^\nu_{\{1\}}-\partial^\nu J^\mu_{\{1\}}\big).$ In the right-hand side of this equation we add the identically zero term $-[1/(\chi c)]\partial_\mu\varepsilon^{\mu\nu\kappa\lambda}\partial_\kappa J_\lambda^{\{2\}}\equiv 0$ and as a final result we obtain equation (80).
We can apply the dual transformations equations (79) to (80) and obtain equation (81)

Consider the Green function $D=D(x,x')$ for the four-dimensional wave equation \cite{23}: $\partial_\mu\partial^\mu D=\delta^{(4)}(x-x'),$
where $\delta^{(4)}(x\!-\!x')$ is the four-dimensional delta function. The function $D$ satisfies the property $\partial^\mu D=-\partial'^\mu D.$ In appendix B we show the following results:
\begin{eqnarray}
\quad\int\! d^4x' D\,\partial'_\mu {\cal F}^{\mu\alpha\,.\,.\,.}=\partial_\mu\! \!\int\! d^4x'\, D\, {\cal F}^{\mu\alpha \,.\,.\,},\\
\int\! d^4x' D\,\partial'_\mu \partial'^\mu {\cal F}^{\alpha\beta \,.\,.\,.}=\int\! d^4x'{\cal F}^{\alpha\beta\,.\,.\,.}\partial'_\mu \partial'^\mu\, D \nonumber\\ \qquad\qquad\qquad\quad\;={\cal F}^{\alpha\beta...}(x),
\end{eqnarray}
where the integrals are taken over all space-time and ${\cal F}^{\mu\alpha\,.\,.\,.(x)}$ is a Lorentz tensor.

We evaluate equation (80) at $x'$, multiply by $D$ and integrate over all space-time,
\begin{eqnarray}
\int\! d^4x' D\,\partial'_\mu\partial'^\mu J^\nu_{\{1\}}&= \int\! d^4x'\, D\,\partial'_\mu\Bigg(\partial'^\mu J^\nu_{\{1\}}-\partial'^\nu J^\mu_{\{1\}}-\frac{1}{\chi c} \varepsilon^{\mu\nu\kappa\lambda}\partial'_\kappa J_\lambda^{\{2\}}\Bigg).
\end{eqnarray}
Using equations (82) and (83) in equation (84), we obtain equation (70). As already noted, we apply the changes given in equation (79) to equation (70) and obtain equation (71). Once equations (70) and (71) have been derived, the four-current theorem is demonstrated. The proof of its corollary follows immediately  by substituting  equations (72) and (73) into equations (70) and (71).  We note that equations (70) and (71) can also be justified by applying $\partial_\alpha\partial^{\alpha}$ to them and showing that two identities result. In fact, if we apply $\partial_\alpha\partial^{\alpha}$ to equations (70) and (71) and perform the corresponding integration over spacetime then we obtain the identities given in equations (80) and (81).

Integration of the wave equation $\partial_\mu\partial^\mu D=\delta^{(4)}(x-x')$ yields two physically-different solutions \cite{23}:
The retarded solution $D_r(x,x')\!=\!\Theta(x_0\!-\!x'_0)\delta[(x\!-\!x')^2]/(2\pi)$ and the advanced solution $D_a(x,x')\!=\!\Theta(x_0'\!-\!x_0)\delta[(x\!-\!x')^2]/(2\pi)$ where $\Theta$ is the theta function. For physical applications we must identify $D$ with $D_r$. We note that the tensor fields given in equations (72) and (73) and their associated equations given in equation (74) are invariant under the dual transformations in equations (75), (76) and the dual transformations:
\begin{eqnarray}
F'^{\mu\nu}_{\{1\}}&= F^{\mu\nu}_{\{1\}}\cos\theta +F^{\mu\nu}_{\{2\}}\sin \theta, \\
F'^{\mu\nu}_{\{2\}}&= -F^{\mu\nu}_{\{2\}}\sin\theta +F^{\mu\nu}_{\{1\}}\cos \theta.
\end{eqnarray}

To obtain the covariant form of Maxwell's equations with magnetic monopoles, we identify again $c$ with the speed of light in vacuum and assume the values of $\alpha, \beta$ and $\chi$ given in the Table 1. We then make the following identifications:
\begin{eqnarray}
F^{\mu\nu}_{\{1\}}=F^{\mu\nu},\quad F^{\mu\nu}_{\{2\}}=\!^*\!{F}^{\mu\nu},\quad J^\nu_{\{1\}}=\beta J^\nu_{\rm e},\quad J^\nu_{\{2\}}=\beta J^\nu_{\rm m},
\end{eqnarray}
where $F^{\mu\nu}$ is the electromagnetic tensor field and $^*\!{F}^{\mu\nu}=(1/2)\varepsilon^{\mu\nu\kappa\lambda}F_{\kappa\lambda}$ is its associated dual.
The components of these tensors are defined by
\begin{eqnarray}
F^{i 0}=\frac{1}{\chi c}\big(\v E\big)^i,\quad F^{i j}=-\varepsilon^{ijk}\big(\v B\big)_k,\quad ^*\!{F}^{i 0}=\big(\v B\big)^i,\quad ^*\!{F}^{i j}=\frac{1}{\chi c}\varepsilon^{ijk}\big(\v E\big)_k,
\end{eqnarray}
where $(\v E\big)^i$ and $(\v B\big)^i$ denote the components of the electric and magnetic fields. The electric and magnetic four-current densities given in equation (87) are respectively defined by
\begin{eqnarray}
J^\nu_{\rm e}=\big(c\rho_{\rm e},\v J_{\rm e}\big), \quad J^\nu_{\rm m}=\big(c\rho_{\rm m}, \v J_{\rm m}\big),
\end{eqnarray}
which are localized in space-time and satisfy the continuity equations $\partial_\nu J^\nu_{\rm e} =0$ and $\partial_\nu J^\nu_{\rm m} =0$. When the identifications given in equation (87) are used in equations (72) and (73), we get the electromagnetic tensor field and its dual:
\begin{eqnarray}
F^{\mu\nu}&=\beta\!\int\! d^4x' D\,\Bigg(\partial'^\mu J^\nu_{\rm e}-\partial'^\nu J^\mu_{\rm e}-\frac{1}{\chi
 c}\varepsilon^{\mu\nu\kappa\lambda}\partial'_\kappa J_\lambda^{\rm m}\Bigg),\\
^*\!{F}^{\mu\nu}&=\beta\!\int\! d^4x' D\,\Bigg( \frac{1}{\chi c}\big(\partial'^\mu J^\nu_{\rm m}-\partial'^\nu J^\mu_{\rm m}\big)+ \varepsilon^{\mu\nu\kappa\lambda}\partial'_\kappa J_\lambda^{\rm e}\Bigg).
\end{eqnarray}
Analogously, when the identifications given in equation (87) are used in the field equations displayed in equation (74) we obtain the corresponding field equations
\begin{eqnarray}
\partial_\mu F^{\mu\nu}=\beta J^\nu_{\rm e},\quad \partial_\mu\,\!^*\!{F}^{\mu\nu}=\frac{\beta}{\chi c}J^\nu_{\rm m},
\end{eqnarray}
which are the covariant form of Maxwell's equations with magnetic charges \cite{27}.

Using the identifications given in (88) we can construct the following four-vectors \cite{27}:
\begin{eqnarray}
\partial_\mu F^{\mu\nu}&=\Bigg(\frac{1}{\chi c}\nabla\cdot \v E, \nabla\times \v B-\frac{1}{\chi c^2}\frac{\partial \v E}{\partial t}\Bigg),\\
\partial_\mu\,\!^*\!{F}^{\mu\nu}&=\Bigg(\nabla\cdot \v B, -\frac{1}{\chi c}\nabla\times \v E-\frac{1}{c}\frac{\partial \v E}{\partial t}\Bigg).
\end{eqnarray}
The three-dimensional form of Maxwell's equations with magnetic monopoles appearing in equations (36)-(39) can be directly obtained using
the equations given in (89) and (92)-(94).

In a second application of the corollary associated with the covariant form of the theorem, we can obtain the covariant form of Maxwell's equations in material media.
We again identify $c$ with the speed of light in vacuum and assume that $\alpha, \beta$ and $\chi$ are given in Table 1. We make the identifications
\begin{eqnarray}
F^{\mu\nu}_{\{1\}}=F^{\mu\nu}, \quad F^{\mu\nu}_{\{2\}}=\!^*\!{F}^{\mu\nu},\quad J^\nu_{\{1\}}=\beta J^\nu_{\rm e}-\partial_\mu Q^{\mu\nu}, \quad J^\nu_{\{2\}}=0,
\end{eqnarray}
where $F^{\mu\nu}$ is again the electromagnetic field tensor and $Q^{\mu\nu}$ is the polarization-magnetization tensor whose components are defined by
\begin{eqnarray}
Q^{i 0}=\frac{\alpha}{\chi c}\big(\v P\big)^i,\quad Q^{i j}=\frac{\beta}{\chi}\varepsilon^{ijk}\big(\v M\big)_k.
\end{eqnarray}
 Using the identifications given in equation (95) we can show that equation (72) becomes the electromagnetic field tensor produced by free and bound sources and equation
 (73) becomes its associated dual tensor:
\begin{eqnarray}
F^{\mu\nu}&=\beta\!\int\! d^4x' D\,\Bigg[\partial'^\mu\bigg(J^\nu_{\rm e}-\frac{1}{\beta}\partial'_\sigma Q^{\sigma\nu}\bigg)-\partial'^\nu
\bigg(J^\mu_{\rm e}-\frac{1}{\beta}\partial'_\sigma Q^{\sigma\mu}\bigg)\Bigg],\\
^*\!{F}^{\mu\nu}&=\beta\!\int\! d^4x' D\,\Bigg[\frac{1}{\chi c} \varepsilon^{\mu\nu\kappa\lambda}\partial'_\kappa \bigg(J_\lambda^{\rm e}-\frac{1}{\beta}\partial'^\sigma Q_{\sigma\lambda}\bigg)\Bigg],
\end{eqnarray}
and the field equations appearing in equation (74) become the covariant for of Maxwell's equations in material media:
\begin{eqnarray}
\partial_\mu F^{\mu\nu}=\beta J^\nu_{\rm e}-\partial_\mu Q^{\mu\nu},\quad \partial_\mu\,\!^*\!{F}^{\mu\nu}=0.
\end{eqnarray}

Using equations (93), (94), (99)  and the four-vector
\begin{eqnarray}
\beta J^\nu_{\rm e}-\partial_\mu Q^{\mu\nu} =\beta\,\bigg(c\rho_{\rm e} - c\nabla\cdot\v P, \v J_e +\frac{1}{\chi}\nabla\times \v M+\frac{\partial \v P}{\partial t}\bigg),
\end{eqnarray}
we obtain the three-dimensional form of Maxwell's equations given in equations (61)-(64). Notice that the field equations given in equation (99) can alternatively be written as
\begin{eqnarray}
\partial_\mu U^{\mu\nu}=\beta J^\nu_{\rm e},\quad \partial_\mu\,\!^*\!{F}^{\mu\nu}=0,
\end{eqnarray}
where the components of the tensor field $U^{\mu\nu}$ are defined as
\begin{eqnarray}
U^{i 0}&=\frac{1}{k_1\chi c}\big(\v D\big)^i\equiv\frac{1}{\chi c}(\v E)^i+\frac{\alpha}{\chi c}(\v P)^i=F^{i 0}+Q^{i 0}, \\
U^{i j}&=-\frac{1}{k_2}\varepsilon^{ijk}\big(\v H\big)_k\equiv -\varepsilon^{ijk}\big(\v B\big)_k+\frac{\beta}{\chi}\varepsilon^{ijk}\big(\v M\big)_k=F^{i j}+Q^{i j}.
\end{eqnarray}
From these equations it follows $\big(\v D\big)^i \equiv k_1\big [\big(\v E\big)^i+\alpha\big(\v P\big)^i\big]$ and $
\big(\v H\big)^i \equiv k_2\big[\big(\v B\big)^i-(\beta/\chi)\big(\v M\big)^i\big]$, i.e., the components of the three-dimensional equations given in equations (65) and (66).
\vskip 10pt
\textcolor[rgb]{0.00,0.00,0.75}{\section{Discussion}}
\noindent Can Maxwell's equations be axiomatized? Let us respond this question for the particular case of  Maxwell's standard equations with only electric sources in vacuum. We first write
 \begin{eqnarray}
\v F_1=\v E/\alpha,\; \v F_2=\chi \v B/\beta, \;\rho_1=\rho_{\rm e},\; \rho_2=0, \;\v J_1=\v J_{\rm e}\; \v J_2=0
 \end{eqnarray}
in equations (9)-(12). We then identify $\rho_{\rm e}$ and $\v J_{\rm e}$ with the electric charge and current densities and assume that they
are localized in space and time and satisfy the continuity equation: $\nabla\cdot\v J_{\rm e}+ \partial\rho_{\rm e}/\partial t=0.$ If $c$ is the speed of light in vacuum and the values of $\alpha, \beta$ and $\chi$ are specified in the table 1 then we obtain the Maxwell equations displayed in equations (46)-(49). The basic ingredients underlying equations (46)-(49) are then the continuity equation for the electric charge and the Green function of the wave equation. In order to satisfy causality we choose the retarded form of this Green function. Accordingly, the principle of electric charge conservation (represented by the continuity equation) and the causality principle (represented by the retarded Green function of the wave equation) are the most fundamental physical axioms that underlie Maxwell's equations. On the basis on the theorem formulated here, the answer to our initial question is in the affirmative. Charge conservation and causality are the basic postulates of Maxwell's equations.

While the first postulate (charge conservation) is a necessary but not a sufficient condition for Maxwell's standard equations, the second postulate (causality) is a sufficient but not a necessary condition for these equations. Let us explain these statements. Using the continuity equation we can obtain the identity
\begin{eqnarray}
\nabla^{2}\v J= -\nabla\times \big(\nabla\times\v J\big)+ \frac{\partial}{\partial t}\big(-\nabla\rho\big).
 \end{eqnarray}
 If we consider the additional identity
 \begin{eqnarray}
 \nabla^{2}\rho= -\nabla\cdot\big(-\!\nabla\rho\big),
  \end{eqnarray}
then the set formed by equations (105) and (106) is different from the set formed by equations (22) and (20).  Using the infinity-space instantaneous Green function \cite{31}: $G_I(\v x,t;\v x',t')\!= \!\delta(t'\!-\!t)/{R}$, which satisfies the Poisson equation $\nabla^{2}G_I\!= \!-4\pi\delta\big(\v x\!-\!\v x'\big)\delta(t\!-\!t')$ [see section 7], as an integrating factor for equations (105) and (106), we can formulate another theorem for the functions $\rho$ and $\v J$, whose corollary yields the time-dependent electric and magnetic fields: $\v E(\v x,t)\!=\!-\alpha\int d^3x'\nabla'\rho(\v x',t)/(4\pi)$ and $\v B(\v x,t)\!=\!\beta\int d^3x'\nabla'\times\v J(\v x',t)/(4\pi)$ and their associated field equations: $\nabla\cdot\v E=\alpha\rho, \nabla\times\v B-(\beta/\alpha)\partial\v E/\partial t=\beta \v J, \nabla\cdot\v B=0$ and $\nabla\times \v E=0$. These equations describe the fields of an instantaneous electrodynamics \cite{28}. Therefore, the continuity equation together with another Green's function distinct to the Green function of the wave equation (for example, the Green function $G_I$) can lead to another theory different from that of Maxwell. This example shows that charge conservation is a necessary but not a sufficient condition to build Maxwell's equations. As is well-known, electric charge conservation
has a strong experimental support \cite{29}. Conceivably, the electron would decay into a photon and a neutrino if the law of electric
charge conservation is violated. A recent test of electric charge conservation with borexino \cite{30} suggests that the time-decay of the electron is
$\tau \geq 6.6\times 10^{28}$ yr. This result strongly supports the validity of charge conservation.

As already pointed out, the Green function appearing in equations (3)-(6) can be identified either with its retarded form or  with its advanced form.
Consequently, the retarded Green function of the wave equation is not a necessary condition to construct Maxwell's equations because we can equally choose the advanced Green function. Causality is shown to be a sufficient but not a necessary condition for Maxwell's equations.

The extension of the above results to the more general case of Maxwell's equations with magnetic monopoles is straightforward. The fundamental physical axioms behind these generalized equations are the principles of conservation of electric and magnetic charges (represented by their associated continuity equations) and the principle of causality (represented by the infinite-space retarded Green function of the wave equation).

In the derivation of Maxwell's equations presented here the Lorentz symmetry does not play the role of a fundamental physical axiom for these equations like do charge conservation and causality. In section 3 we have derived Maxwell's equations in the three-dimensional space without any explicit reference to the Lorentz symmetry.

 To finish this section we note that the corollary expressed in equations (7)-(12) is of general validity and can also be applied to other field theories. For example, it can be used to find a set of gravitational field equations. Let us specialize in SI units and write
 \begin{eqnarray}
\!\!\!\!\!\!\!\!\!\!\!\!\!\!\!\!\!\!\chi=1, \;\v F_1=\v g,\; \v F_2=\v k c^2,\; \rho_1=-4\pi\mathfrak{S}\rho_{\textsc{m}},\; \rho_2=0,\; \v J_1=-4\pi\mathfrak{S}\v J_{\textsc{m}},\; \v J_2=0,
\end{eqnarray}
where $\rho_{\textsc{m}}$ and $\v J_{\textsc{m}}$ are the mass density and the mass current density. These densities are assumed to satisfy the continuity equation $\nabla\cdot\v J_{\textsc{m}}+\partial\rho_{\textsc{m}}/\partial t=0$. The time-dependent vector fields $\v g$ and $\v k$ are respectively the gravitoelectric and gravitomagnetic fields with $\mathfrak{S}$ being the universal gravitational constant. Using the identifications given in equation (107) we can see that equations (7) and (8) yield
\begin{eqnarray}
\v g &= \mathfrak{S}\!\iint \!\!d^3x'dt'G\Bigg(\nabla'\!\rho_{\textsc{m}}\!+\!\frac{1}{c^2}\frac{\partial\v J_{\textsc{m}} }{\partial t'}\Bigg),\\
\v K &= -\frac{\mathfrak{S}}{c^2}\!\iint \!\!d^3x'dt'G\nabla'\times\v J_{\textsc{m}},
\end{eqnarray}
and equations (9)-(12) become
\begin{eqnarray}
\nabla\cdot\v g= -4\pi\mathfrak{S}\rho_{\textsc{m}},\\
\nabla\cdot\v k=0,\\
\nabla\times \v k-\frac{1}{c^2}\frac{\partial\v g}{\partial t}=-\frac{4\pi\mathfrak{S} }{c^2}\v J_{\textsc{m}},\\
\nabla\times \v g + \frac{\partial\v k}{\partial t}=0.
\end{eqnarray}
If we define the gravitational permittivity constant of vacuum as $\tilde{\epsilon}_0=1/(4\pi\mathfrak{S})$ and the gravitational permeability constant of vacuum as $\tilde{\mu}_0=4\pi\mathfrak{S}/c^2$ then $\tilde{\epsilon}_0\tilde{\mu}_0=1/c^2$. Therefore equations (110) and (112) take a form resembling
the Gauss and Amper\'{e}-Maxwell laws:
\begin{eqnarray}
\nabla\cdot\v g=-\rho_{\textsc{m}}/\tilde{\epsilon}_0,\\
\nabla\times \v k-\tilde{\epsilon}_0\tilde{\mu}_0\frac{\partial\v g}{\partial t}=-\tilde{\mu}_0\v J_{\textsc{m}}.
\end{eqnarray}
The existence of the gravitomagnetic vector field was first suggested by Heaviside \cite{20} who essentially introduced equations (110)-(113).
 Jefimenko \cite{21} and McDonald \cite{22} have also discussed equations (110)-(113).
\vskip 10pt

\textcolor[rgb]{0.00,0.00,0.75}{\section{Pedagogical ingredient: Elementary introduction to the retarded Green function }}

\noindent The Green's function of the wave equation is not typically discussed in undergraduate books of electromagnetism. The introduction of this function is usually delayed until the graduate level and therefore the axiomatic approach to Maxwell's equations suggested here, which makes an extensive use of the infinity-space retarded Green function of the wave equation, turns out to be suitable for a graduate course of electrodynamics. However, we have suggested elsewhere \cite {31} that this retarded Green function could be introduced early by means of a non-rigourous and heuristic procedure at undergraduate level. Accordingly, the three-dimensional axiomatic derivation proposed in this paper could be suitable for an advanced undergraduate course of electrodynamics. For completeness of this paper, we are going to briefly discuss this informal treatment of the retarded Green function of the wave equation.

As is well-known the function $1/|\v x\!- \!\v x'|\!\equiv\! 1/R$ satisfies the identity $\nabla^2(1/R)\!=\!-4\pi\delta(\v x\!-\!\v x') $. If we multiply this last equation by $\delta(t-t')$ then we obtain
\begin{equation}
\nabla^2\bigg[\frac{\delta(t-t')}{R}\bigg]=-4\pi\delta(\v x-\v x')\delta(t-t').
\end{equation}
We define the infinity-space Green function $G_I\equiv G_I(\v x,t;\v x',t')$ of the Poisson equation by
\begin{equation}
G_I=\frac{\delta(t\!-\!t')}{R}.
\end{equation}
This means that the solution of the Poisson equation
\begin{equation}
\nabla^2F(\v x,t)=-4\pi f(\v x,t).
\end{equation}
can be written as
\begin{equation}
F(\v x,t)=\iint \!\!d^3x'dt'G_I f(\v x',t').
\end{equation}
According to this solution the value of the field $F$ at the point $\v x$ at the time $t$ is related to the values of the source $F$ at the point $\v x'$ at the instant time $t'$ in such a way that $t=t'$, i.e., after time integration equation (119) becomes
\begin{equation}
F(\v x,t)=\int \!\!d^3x'\frac{f(\v x',t)}{R}.
\end{equation}
The function $G_I$ can be called the instantaneous form of the Green function for the Poisson equation because it exhibits an acausal behavior. The argument of the delta function within $G_I$ shows that an effect observed at the point $\v x$ at the time $t$ is connected by the action of a source a distance $R$ away at the point $\v x'$ at the same instant of time $t'=t$. Stated in other words: in an instantaneous interaction the time of the field is the same as the time of its source: $t'=t$.

But instantaneous interactions violate the principle of causality which states that causes (sources) precede in time to their effects (fields), i.e., the time of the source is previous to the time of the field: $t'\!<\! t$. Therefore, causality demands to modify the Green function $G_I$ in such a way that the modified Green function $G$ exhibits a retarded behavior. Accordingly, the argument of the delta function within $G$ must show that an effect observed at the point $\v x$ at the time $t$ is caused by the action of a source a distance $R$ away at the point $\v x'$ at the
retarded time $t'\!=\!t\! -t_0$, where $t_0\!>\!0$ is the time required for the carrier of the interaction to travel the distance $R$ between the point $\v x'$ and the point $\v x$. The carrier of the electromagnetic interaction is the photon which moves with the speed of light $c$ in vacuum, which implies $t_0\!=\!R/c$ and consequently the retarded time takes the form  $t'\!=t\! -\!R/c$. Put differently, causality demands the change $\delta(t\!-\!t')\to \delta(t\!-\!t' \!+\!R/c)$, or equivalently, the change
\begin{equation}
G_I=\frac{\delta(t-\!t')}{R}\longrightarrow G=\frac{\delta(t'\!-\!t+R/c)}{R},
\end{equation}
 where we have used the property $\delta(u)=\delta(-u)$. The final step is to verify that the infinite-space retarded Green function satisfies the wave equation
 \begin{equation}
\nabla^2\bigg[\frac{\delta(t'\!-\!t+R/c)}{R}\bigg]-\frac{1}{c^2}\frac{\partial^2}{\partial t^2}\bigg[\frac{\delta(t'\!-\!t+R/c)}{R}\bigg]=-4\pi\delta(\v x-\v x')\delta(t-t').
\end{equation}
Let us write $u\!=t'\!-\!t+R/c$. Thus $G\!=\!\delta(u)/R.$ Now we calculate the Laplacian of $G$:
  \begin{equation}
\nabla^2\bigg[\frac{\delta(u)}{R}\bigg]= \delta(u)\nabla^2\bigg(\frac{1}{R}\bigg) +2\nabla\delta(u)\cdot\nabla\bigg[\frac{1}{R}\bigg]+ \frac{1}{R}\nabla^2\delta(u).
\end{equation}
 Using the following results \cite{18}:
\begin{equation}
\!\!\!\!\!\!\!\!\!\!\!\!\!\!\!\!\!\!\!\!\!\!\!\!\!\!\!\!\!\!\!\!\!\!\!\!\!\!\!\!\!
\nabla^2\bigg[\!\frac{1}{R}\!\bigg]\!=\!-4\pi\delta(\v x\!-\!\v x'),\;\nabla\delta(u)\!=\!-\frac{\hat{\v R}}{c}\frac{\partial \delta(u)}{\partial t}, \;
\nabla\bigg[\!\frac{1}{R}\!\bigg]\!=\!-\frac{\hat{\v R}}{R^2},\; \nabla^2\delta(u)\!=\!\frac{1}{c^2}\frac{\partial^2 \delta(u)}{\partial t^2}-\frac{2}{Rc}\frac{\partial \delta(u)}{\partial t},
\end{equation}
where $\hat{\v R}=\v R/R\equiv (\v x-\v x')/|\v x-\v x'|$, equation (123) becomes
\begin{equation}
\!\!\!\!\!\!\!\!\!\!\!\!\!\!\!\!\!\!\!\!\!\nabla^2\bigg[\frac{\delta(t'\!-\!t+R/c)}{R}\bigg]=-4\pi\delta(\v x\!-\!\v x')\delta(t\!-t'\!-R/c) +\frac{1}{c^2R}\frac{\partial^2}{\partial t^2}\delta(t'\!-\!t+R/c),
\end{equation}
which is essentially equation (122). A more formal verification that $G\!=\!\delta(u)/R$ satisfies equation (122) has been given by Rowe \cite{32}.

\textcolor[rgb]{0.00,0.00,0.75}{\section{Conclusion}}
\noindent The idea of looking for an axiomatization of physics was suggested by Hilbert \cite{33} in 1900. The 6th problem of his celebrated list of problems consists in treating ``...by means of axioms those physical sciences in which mathematics plays an important part.'' A satisfactory answer to this problem was given by Einstein in his special relativity. Partial answers to the  Hilbert's 6th problem have been suggested in the cases of quantum mechanics and general relativity (see for example references \cite{1} and \cite{2}). But in the case of classical electrodynamics, a branch of physics where mathematics is extremely important, a convincing axiomatization of Maxwell's equations is still pending (see reference \cite{3} for an attempt).

Here we have suggested an axiomatization of Maxwell's equations including magnetic charges as well as electric ones. Our starting point was
a theorem formulated for two sets of functions localized in space and time. If each set satisfies a continuity equation then the theorem provides an integro-differential representation for the involved functions in terms of their derivatives and the retarded Green function of the wave equation. A corollary of this theorem led us to Maxwell's equations with electric and magnetic charges.
We have then concluded that the principles of conservation of electric and magnetic charges and the principle of causality are the basic postulates underlying Maxwell's generalized equations. We have also expressed our results in a manifestly covariant form.

For the standard case of Maxwell's equations with only electric charges, we have again emphasized \cite{18, 19} that causality and charge conservation are the most fundamental axioms underlying these equations. {\it Finally, it all boils down to saying that charges are conserved and produce causal effects. This is the essential content of Maxwell's equations.}
\vskip 10pt
\textcolor[rgb]{0.00,0.00,0.75}{\section*{Acknowledgement}}

\noindent I dedicate this paper to the memory of Professor John David Jackson.\dag \; The grey, red and blue editions of his {\it Classical Electrodynamics} increased my fascination for Maxwell's equations.

\vskip 10pt
\appendix

\textcolor[rgb]{0.00,0.00,0.75}{\section{Derivation of equations (26)-(28) }}
\noindent Consider the localized quantity ${\cal F}\!=\!{\cal F}(\v x',t')$ and the infinity-space Green function of the wave equation  $G\!=\!G(\v x,t;\v x',t')$. Using the identities $\nabla'\diamond (G{\cal F})\!=\! G\nabla'\diamond {\cal F}+ {\cal F}\diamond \nabla'G$ and $\nabla\diamond (G{\cal F})={\cal F}\diamond \nabla G$ and the property
$\nabla G =-\nabla' G$, we can obtain the relation
\begin{eqnarray}
G\nabla'\diamond {\cal F} =\nabla\diamond (G{\cal F})+\nabla'\diamond (G{\cal F}).
\end{eqnarray}
We integrate equation (A.1) over space and time,
\begin{eqnarray}
\!\!\!\!\!\!\!\!\!\!\!\!\!\!\!\iint \!\!d^3x'dt'G\nabla'\diamond {\cal F}=\iint \!\!d^3x'dt'\nabla\diamond(G {\cal F})+\int\!dt'\Bigg[\!\int d^3x'\nabla'\diamond(G{\cal F})\Bigg].
\end{eqnarray}
We extract the operator $\nabla\diamond$ from the first integral on the right-hand side and transform the volume integral in the second term into a surface integral
\begin{eqnarray}
\iint \!\!d^3x'dt'G\nabla'\diamond {\cal F}=\nabla\diamond\iint \!\!d^3x'dt'G {\cal F}+\int\!dt'\Bigg[\oint d\v S'\diamond(G{\cal F})\Bigg].
\end{eqnarray}
The surface integral in equation (A.3) vanishes at spatial infinity because ${\cal F}$ is assumed to be spatially localized. Therefore equation (A.3) reduces to equation (26). Similarly, the identities $\partial(G{\cal F})/\partial t'\!=\! (\partial G/\partial t'){\cal F}+ G(\partial {\cal F}/\partial t')$ and $\partial(G{\cal F})/\partial t\!=\!(\partial G/\partial t'){\cal F}$ and the property $\partial G/\partial t=-\partial G/\partial t'$ yield the relation
\begin{eqnarray}
G\frac{\partial {\cal F}}{\partial t'}=\frac{\partial(G{\cal F})}{\partial t}+\frac{\partial(G{\cal F})}{\partial t'}.
\end{eqnarray}
We integrate this equation over space and time,
\begin{eqnarray}
\iint \!\!d^3x'dt'G\frac{\partial {\cal F}}{\partial t'}=\iint \!\!d^3x'dt'\frac{\partial(G {\cal F})}{\partial t}+\int \!d^3x' \Bigg[\int dt' \frac{\partial(G{\cal F})}{\partial t'}\Bigg].
\end{eqnarray}
We extract the operator $\partial/\partial t$ from the first term on the right-hand side and perform the time integration in the second term, obtaining
\begin{eqnarray}
\iint \!\!d^3x'dt'G\frac{\partial {\cal F}}{\partial t'}=\frac{\partial}{\partial t}\iint\!\! d^3x'dt' G {\cal F} +\int \!d^3x\big[G {\cal F}\big]_{-\infty}^{+\infty}.
\end{eqnarray}
The second term on the right-hand side vanishes because  ${\cal F}$ is bounded in time and $G=0$ at $t'\pm \infty$. Thus equation (A.6) reduces to equation (27). By a direct calculation we can verify the following identity
\begin{eqnarray}
\!\!\!\!\!\!\!\!\!\!\!\!\!\!\!\!\!\!\!\!\!\frac{\partial}{\partial t'}\Bigg(\frac{{\cal F}}{c^2}\frac{\partial G}{\partial t'}-\frac{G}{c^2}\frac{\partial{\cal F}}{\partial t'}\Bigg)
+\nabla'\diamond \big(G\nabla'\diamond {\cal F}-{\cal F}\diamond\nabla' G\big)= G\Box'^2{\cal F} -{\cal F}\Box'^2G.
\end{eqnarray}
Time and space integration of this identity leads to
\begin{eqnarray}
\!\!\!\!\!\!\!\!\!\!\!\!\!\!\!\!\!\!\!\!\!\!\iint \!\!d^3x'dt'G\Box'^2{\cal F}&=\iint \!\!d^3x'dt'{\cal F}\Box'^2G +
\!\int\! d^3x'\Bigg[\int\!dt'\frac{\partial}{\partial t'}\Bigg(\frac{{\cal F}}{c^2}\frac{\partial G}{\partial t'}-\frac{G}{c^2}\frac{\partial{\cal F}}{\partial t'}\Bigg)\Bigg]\nonumber\\
&\;\;\;+\int\!dt'\Bigg[\int\!d^3x'\nabla'\diamond \big(G\nabla'\diamond {\cal F}-{\cal F}\diamond\nabla' G\big)\Bigg]\nonumber\\
&=\!\iint \!\!d^3x'dt'{\cal F}\Box'^2G\! +\!
\!\int\!\! d^3x'\Bigg[\frac{{\cal F}}{c^2}\frac{\partial G}{\partial t'}\!-\!\frac{G}{c^2}\frac{\partial{\cal F}}{\partial t'}\Bigg]_{-\infty}^{+\infty}
\nonumber\\
&\;\;\;+\!\int\!\!dt'\Bigg[\oint d\v S'\diamond\big(G\nabla'\diamond {\cal F}\!-\!{\cal F}\diamond\nabla' G\big)\Bigg].
 \end{eqnarray}
The second term on the right-hand side vanishes because ${\cal F}$ and $\partial{\cal F}/\partial t$ are bounded
in time and both $G=0$ and $\partial G/\partial t'=0$ at $t'\pm \infty$. The third term on the right-hand side vanishes at spatial infinity because ${\cal F}$ and $\nabla' \diamond{\cal F}$ are spatially localized. Therefore, equation (A.8) reduces to the first equality of equation (28). The second equality in equation (28) follows from
substituting $\Box'^{2}G= -4\pi\delta(\v x-\v x')\delta(t-t')$ in the first equality and performing the corresponding integration over all space and all time.

\textcolor[rgb]{0.00,0.00,0.75}{\section{Derivation of equations (82) and (83)}}
Consider the generic tensor ${\cal F}^{\mu\alpha...}(x')$ and the Green function $D\!=\!D(x,x')$. Using the property $\partial^\mu D=-\partial'^\mu D$ and the identities $\partial'_\mu(D{\cal F}^{\mu\alpha...})=\partial'_\mu D{\cal F}^{\mu\alpha...}+  D\partial'_\mu{\cal F}^{\mu\alpha...}$
and $\partial_\mu(D{\cal F}^{\mu\alpha...})=\partial_\mu D{\cal F}^{\mu\alpha...}$, we can derive the identity
\begin{eqnarray}
D\partial'_\mu {\cal F}^{\mu\alpha...}=\partial_\mu \big(D {\cal F}^{\mu\alpha...}\big)+ \partial'_\mu \big(D {\cal F}^{\mu\alpha...}\big).
\end{eqnarray}
Integration of this identity over all space-time reads
\begin{eqnarray}
\int\! d^4x'D\partial'_\mu {\cal F}^{\mu\alpha...}&=\int\! d^4x'\partial_\mu\big(D{\cal F}^{\mu\alpha...}\big)+\int\! d^4x'\partial'_\mu\big(D{\cal F}^{\mu\alpha...}\big).
\end{eqnarray}
We extract $\partial_\mu$ from the first integral and transform the second term into a surface integral
\begin{eqnarray}
\int\! d^4x'D\partial'_\mu {\cal F}^{\mu\alpha...}=\partial_\mu\int\! d^4x'D{\cal F}^{\mu\alpha...}+\oint dS'_\mu D{\cal F}^{\mu\alpha...}.
\end{eqnarray}
The surface integral in equation (B.3) vanishes at the space-time infinity because ${\cal F}^{\mu\alpha...}$ is localized in space-time. Therefore equation (B.3) reduces to equation (82). A direct calculation allows us to show the identity
\begin{eqnarray}
\partial'_\mu\big(\partial'^\mu D{\cal F}^{\alpha\beta...} -D \partial'^\mu{\cal F}^{\alpha\beta...}\big)={\cal F}^{\alpha\beta...}\partial'_\mu\partial'^\mu D- D\partial'_\mu\partial'^\mu{\cal F}^{\alpha\beta...}.
\end{eqnarray}
Integration of equation (B.4) over all space-time yields
\begin{eqnarray}
\!\!\!\!\!\!\!\!\!\!\!\!\!\!\!\!\!\!\!\!\!\!\!\!\!\!\!\!\int\! d^4x'D\partial'_\mu \partial'^\mu {\cal F}^{\alpha\beta...}=\int\! d^4x'{\cal F}^{\alpha\beta...}\partial'_\mu\partial'^\mu D-\int\! d^4x'\partial'_\mu\big(\partial'^\mu D{\cal F}^{\alpha\beta...} -D\partial'^\mu{\cal F}^{\alpha\beta...}\big)\nonumber\\
\qquad\quad=\int\! d^4x'{\cal F}^{\alpha\beta...}\partial'_\mu\partial'^\mu D-\oint dS'_\mu\big(\partial'^\mu D{\cal F}^{\alpha\beta...} -D \partial'^\mu{\cal F}^{\alpha\beta...}\big).
\end{eqnarray}
The surface integral vanishes at the space-time infinity because ${\cal F}^{\alpha\beta...}$ and $\partial'^\mu{\cal F}^{\alpha\beta...}$ are localized in space-time and therefore equation (B.5) reduces to the first equality of equation (83). The second equality in equation (83) follows from
the first one by inserting the equation $\partial'_\mu\partial'^\mu D=\delta^{(4)}(x-x')$ and integrating over all space-time.

\textcolor[rgb]{0.00,0.00,0.75}{{}
\end{document}